\documentstyle[aps,multicol,prl,epsf]{revtex}
\begin{document}
\title{Dynamics of Phase Separation under Shear: A Soluble Model}
\author{N. P. Rapapa and A. J. Bray}
\address{Department of Physics and Astronomy, The University,
Manchester M13 9PL, UK}
\date{\today}
\maketitle

\begin{abstract}
The dynamics of phase separation for a binary fluid subjected to a uniform 
shear are solved exactly for a model in which the order parameter is
generalized to a $n$-component vector and the large-$n$ limit taken. 
Characteristic length scales in directions parallel and perpendicular 
to the flow increase as $(t^5/\ln t)^{1/4}$ and $(t/\ln t)^{1/4}$ 
respectively. The structure factor in the shear-flow plane exhibits two 
parallel ridges as observed in experiment. 
\end{abstract}



\begin{multicols}{2}
The dynamics of phase separation under shear has attracted considerable 
theoretical \cite{OnukiRev}, 
experimental \cite{Beysens,Krall,Hashimoto,Lauger} and 
simulational \cite{Rothman,Padilla,Yeomans} 
attention in recent years. In the absence of 
shear, the dynamics of phase separation is now quite well 
understood \cite{Bray}. 
Domains of the two equilibrium phases are formed, and coarsen with time in 
a manner well-described by a dynamical scaling phenomenology with a single 
growing length scale $L(t)$ which generally grows as a power law in time, 
$L(t) \sim t^a$. The structure factor is spherically symmetric, with a 
maximum at wavevector $k_m  \sim L^{-1}$. For binary fluids, the exponent 
$a$ takes different values depending on the dynamical regime under study. 
In order of increasing time, there are `diffusive' ($a=1/3$), 
`viscous hydrodynamic' ($a=1$) and the `inertial hydrodynamic' ($a=2/3$) 
regimes \cite{Bray,Siggia,Furukawa}. The crossover between these regimes 
is determined by the fluid properties (viscosity, density). Here we will 
focus on the diffusive regime, in which hydrodynamic effects can be 
neglected. In the absence of shear, phase separation is described by the 
Cahn-Hilliard equation for the order-parameter field $\phi({\bf r},t)$, 
namely \cite{Bray} 
$\partial_t \phi = -\nabla^2(\nabla^2\phi + \phi - \phi^3)$.

If a uniform shear flow is imposed in the $x$-direction, with shear direction 
$y$, the flow velocity is $v_x = \gamma y$, where $\gamma$ is the shear rate. 
For an incompressible fluid, the Cahn-Hilliard equation acquires an advective 
term ${\bf v}\cdot\nabla\phi = \gamma y \partial_x \phi$ on the left-hand
side. Generalizing to an $n$-component vector order parameter, this gives 
\begin{equation}
\partial_t \vec{\phi} + \gamma y\partial_x\vec{\phi} 
= -\nabla^2\left(\nabla^2\vec{\phi} + \vec{\phi} 
- \frac{1}{n}(\vec\phi)^2\vec{\phi}\right).
\label{CH}
\end{equation}
For a critical mixture quenched into the two-phase region from the 
homogeneous phase, an appropriate initial condition is a Gaussian 
random field with zero mean and short-range correlations: 
$\langle\phi_i({\bf r},0)\phi_j({\bf r}',0)\rangle = 
\Delta\delta_{ij}\delta({\bf r}-{\bf r}')$.

In this Letter we present the first exact solution for a system 
phase-separating under shear, by solving (\ref{CH}) in the limit 
$n \to \infty$. We obtain characteristic length scales 
$L_x \sim k_{mx}^{-1} \sim (t^5/\ln t)^{1/4}$, 
$L_y \sim k_{my}^{-1} \sim (t/\ln t)^{1/4}$, and 
$L_z \sim k_{mz}^{-1} \sim (t/\ln t)^{1/4}$. 
These are extracted from the structure factor, $S({\bf k},t)$ which has four 
maxima, located at ${\bf k} = \pm (k_{mx},-k_{my},\pm k_{mz})$. Thus the 
rotational symmetry of the zero-shear limit is completely broken. 
In the $k_z=0$ plane there are two maxima terminating two long parallel 
ridges as observed in experiment \cite{Beysens}. The structure factor 
exhibits multiscaling \cite{CZ}, rather than simple scaling, but this is 
presumably an artifact of the large-$n$ limit, as in the zero-shear 
case \cite{BH}. At the end of the paper we conjecture how the results 
will be modified for the scalar order parameter appropriate to binary 
mixtures.

In the limit $n \to \infty$, one can replace $(\vec{\phi})^2/n$ in (\ref{CH}) 
by its mean in the usual way, leading to a self-consistent linear equation. 
After Fourier transformation this reads
\begin{equation}
\frac{\partial\phi_{\bf k}}{\partial t} 
- \gamma k_x\frac{\partial \phi_{\bf k}}{\partial k_y} 
= -{\bf k^2}[{\bf k^2} - a(t)]\phi_{\bf k},
\label{CHbign}
\end{equation}
where $\phi$ is (any) one component of $\vec\phi$, and 
$a(t) = 1 - \langle \phi^2\rangle$.

The same equation was studied in a recent Letter by 
Corberi et al.\ \cite{Corberi}, where it was regarded as a 
`self-consistent one-loop' approximation to the scalar 
version of (\ref{CH}). These authors, however, did not solve the equation 
analytically, but rather integrated it numerically (in two space dimensions). 
As a result they were unable to access the asymptotic ($t \to \infty$) 
behavior which is the main focus of the present work. In particular we do 
not find the oscillatory large-time behavior that Corberi et al.\ 
conjectured from their numerical results.

Equation (\ref{CHbign}) can be solved via the change of variables 
$(k_x,k_y,k_z,t) \to (k_x,\sigma,k_z,\tau)$, where $\tau = t$ and 
$\sigma = k_y + \gamma k_xt$. The left-hand side of (\ref{CHbign}) then 
becomes $\partial\phi_{\bf k}/\partial \tau$, and the equation can be 
integrated directly to give, after transforming back to the original 
variables, $\phi_{\bf k}(t) = \phi_{\bf k}(0)\exp f({\bf k},t)$, where 
\begin{eqnarray}
f({\bf k},t) & = & -(k_x^2 + k_z^2)^2t 
- \frac{2(k_x^2 + k_z^2)}{3\gamma k_x}\left((k_y+\gamma k_xt)^3 - k_y^3\right) 
\nonumber \\
&&-\frac{1}{5\gamma k_x}\left((k_y+\gamma k_xt)^5 - k_y^5\right)\nonumber \\
&& + \left(k_x^2 + k_z^2 + (k_y + \gamma k_xt)^2\right)b(t) \nonumber \\
&& - 2\gamma k_x(k_y + \gamma k_xt)c(t) + \gamma^2 k_x^2 e(t),
\label{solution}
\end{eqnarray}
with $b(t) = \int_0^t dt'\,a(t')$, $c(t) = \int_0^t dt'\,t'a(t')$, and 
$e(t) = \int_0^t dt'\,t'^2 a(t')$.

The next step is to determine $a(t)$ self-consistently, using its definition:  
$a(t) = 1 - \langle \phi^2 \rangle = 1 - \sum_{\bf k} S({\bf k},t)$, where 
$S({\bf k},t) = \langle \phi_{\bf k}(t)\phi_{\bf -k}(t) \rangle$ is the 
structure factor. The momentum sum can be evaluated for large $t$ using 
the method of steepest descents. To simplify the analysis we make the 
following ansatz, which will be justified {\em a posteriori}. Naive power 
counting applied to (\ref{CHbign}) suggests characteristic length scales 
$L_y \sim L_z \sim t^{1/4}$, and $L_x \sim t^{5/4}$, where the dominant 
$k_x$-dependence comes from the shear term, and that $a(t) \sim t^{-1/2}$. 
This suggests that only the $k_x$ terms multiplied by $\gamma$ in 
(\ref{solution}) survive in the `scaling' limit. In fact we will find that 
the naive power counting result is modified by logarithms, but the above 
conclusion still holds. Guided by the $\gamma=0$ result, we make the 
ansatz $a(t) \sim (\ln t/t)^{1/2}$ for $t \to \infty$. Then, to leading 
logarithmic accuracy, $b(t) \sim (t \ln t)^{1/2}$, $c(t) \to tb(t)/3$, and 
$e(t) \to t^2b(t)/5$. Inserting these results in (\ref{solution}), and 
making the change of variable 
\begin{equation}
\gamma k_x = \sqrt{\frac{b}{t^3}}\,u,\ \  
k_y = \sqrt{\frac{b}{t}}\,v,\ \ 
k_z = \sqrt{\frac{b}{t}}\,w  
\label{cov}
\end{equation}
gives
\begin{eqnarray}
S({\bf k},t)  & = & \Delta \exp\left(2\frac{b^2}{t}\,F(u,v,w)\right) \\
F(u,v,w) & = & -\frac{1}{5u}\left((u+v)^5-v^5\right) + \frac{8}{15}u^2 
+ \frac{4}{3}uv + v^2 \nonumber \\ 
&& - \frac{2}{3}w^2(u^2+3uv+3v^2) + w^2 - w^4,
\label{F} 
\end{eqnarray}
where contributions to $F$ which vanish as $t \to \infty$ (at fixed $u,v,w$) 
have been dropped.

The self-consistency equation for $a(t)$ reads
\begin{eqnarray}
1 - a(t) & = & \int \frac{d^3k}{(2\pi)^3}\,S({\bf k},t) \nonumber \\
& = &\frac{\Delta b^{3/2}}{(2\pi)^3\gamma t^{5/2}} \nonumber \\
&& \times \int du\,dv\,dw\, \exp\left(2\frac{b^2}{t}\,F(u,v,w)\right). 
\label{self-cons}
\end{eqnarray}
Since $b^2/t$ grows like $\ln t$ by assumption, the triple integral can be 
evaluated by steepest descents for large t. The complete set of stationary 
points $(u,v,w)$ is listed in table 1, together with their type 
(maximum, minimum, saddle point) and the corresponding values of $F$. 
The stationary points with $w=0$ are also stationary points of the 
two-dimensional (2D) theory, and their types in the (u,v) plane are 
listed separately. The $w=0$ structure of $S({\bf k},t)$ is relevant for 
light scattering with the incident beam normal to the $xy$-plane.

The maximum value of $F$, $F_m = 23/60$, occurs at four points in 
$uvw$-space (labeled $f$ in Table 1), corresponding, via (\ref{cov}), to 
four points in momentum space.  
The integral in (\ref{self-cons}) can now be evaluated by steepest descents. 
Since $a(t)$ in (\ref{self-cons}) vanishes (like $(\ln t/t)^{1/2}$) for 
$t \to \infty$ it can be dropped to give
\begin{equation}
1 = {\rm const.}\,\frac{\Delta}{\gamma t b^{3/2}} 
\exp\left(\frac{2F_mb^2}{t}\right),
\label{key}
\end{equation}
whence, to leading logarithmic accuracy, 
\begin{equation}
b(t) = \left(\frac{7}{8F_m}\,t\ln t\right)^{1/2},
\label{b}
\end{equation}
finally justifying our original ansatz. 

\end{multicols}
\medskip

\begin{center}
\begin{tabular}{|c|c|c|c|c|c|c|}
\hline
Label & Position  & Number  & F & Value & Type (3D) & Type (2D) \\ 
\hline
a & $(0,0,0)$ & 1 &   0         & 0 & Min & Min \\ \hline
b & $\pm(2/\sqrt{3},0,0)$ & 2 & 16/45 & .35556 & IS & IMax \\ \hline
c & $\pm(\sqrt{2} - 1/\sqrt{3},-1/\sqrt{2},0)$ &2& $(37 - 12\sqrt{6})/180$ 
& .04225 & S2 & S \\ \hline
d & $\pm(\sqrt{2} + 1/\sqrt{3},-1/\sqrt{2},0)$ &2& $(37 + 12\sqrt{6})/180$ 
& .36885 & S1 & Max \\ \hline 
e & $\pm(0,0,1/\sqrt{2})$ & 2& 1/4 & .25 & S1 & - \\ \hline
f & $\pm(\sqrt{3},-1/\sqrt{3},\pm 1/\sqrt{6})$ &4& 23/60 & .38333 &Max & - \\ 
\hline
\end{tabular}
\end{center}

\underline{Table 1.} Stationary points of $F(u,v,w)$: Max = maximum, 
Min = minimum, S = saddle point (2D), S$n$ = saddle point of type $n$ (the 
matrix of second derivatives has $n$ positive eigenvalues), IS = `inflection 
saddle point' (one positive, one zero, one negative eigenvalue), 
IMax = `inflection maximum' (one zero, one negative eigenvalue). 

\medskip

\begin{multicols}{2}
From equation (\ref{cov}) we can define characteristic length scales for the 
three directions: $L_x = \gamma(t^3/b)^{1/2} \sim \gamma(t^5/\ln t)^{1/4}$, 
and $L_y = L_z = (t/b)^{1/2} \sim (t/\ln t)^{1/4}$, by setting $u=k_xL_x$, 
$v=k_yL_y$, and $w=k_zL_z$. Finally the structure factor is given by 
$S({\bf k},t) = \Delta \exp[(2b^2/t)F(u,v,w)]$. Using (\ref{key}) this can 
be recast as 
\begin{equation}
S({\bf k},t) = {\rm const.}\,(\ln V_s)^{3/2}V_s^{F({\bf q})/F_m},
\label{S(k)}
\end{equation} 
where $V_s(t) = L_xL_yL_z \sim \gamma t^{7/4}/(\ln t)^{3/4}$ is the 
`scale volume' at time $t$, and ${\bf q} = (k_xL_x,k_yL_y,k_zL_z)$ 
is the scaled momentum. 
The corresponding result for zero shear (in space dimension $d$) is 
\cite{Bray,CZ} 
$S({\bf k},t) = {\rm const.}\,(\ln L)^{1/2}L^{d\phi(q)}$, where 
$L = (8t/d\ln t)^{1/4}$, $q=kL$, and $\phi(q) = 2q^2 - q^4$. 
Equation (\ref{S(k)}) does not have a form consistent 
with conventional scaling, $S({\bf k},t) = V_s(t)f({\bf q})$, but 
rather a form of `multiscaling', as in the zero-shear case, where the 
{\em power} of the scaling volume $V_s$ depends on the scaling 
variables. As in the zero-shear case, however, we anticipate that simple 
scaling will be recovered asymptotically for any finite $n$, and that 
the $\ln t$ terms will disappear from the length scales \cite{BH}.

\begin{figure}
\narrowtext
\centerline{\epsfxsize\columnwidth\epsfbox{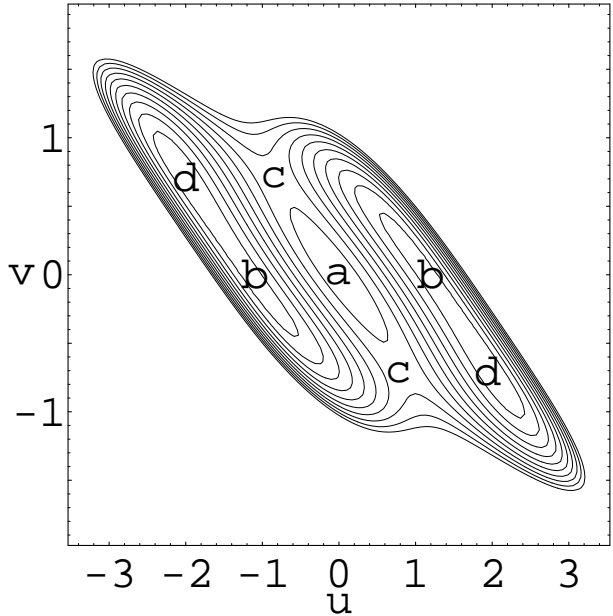}}
\caption{Contour plot of $F(u,v,0)$ showing approximate locations 
of the stationary points $a,b,c,d$ from Table 1. Contour lines for 
$F < -0.1$ are not shown.} 
\label{fuv}
\end{figure}

In measurements of the structure factor by small-angle light scattering,  
the scattering vector ${\bf k}$ is perpendicular to the beam direction. Two 
common experimental arrangements \cite{Beysens} are with the beam in the 
$z$-direction, i.e.\ perpendicular to both the shear and the flow, or in 
the $y$ (i.e. shear) direction. The former corresponds to $k_z=0$, 
the latter to $k_y=0$.
For $k_z=0$, $S(k_x,k_y,0,t)$ is determined by $F(u,v,0)$. In fact, from 
(\ref{S(k)}), $\ln S({\bf k},t) = [F(u,v,0)/F_m]\ln V_s$ 
(plus ${\bf k}$-independent terms), so $F(u,v,0)$ is essentially the 
logarithm of the structure factor divided by $\ln t$. Figure 1 is a 
contour plot of $F(u,v,0)$, showing the stationary points $a,b,c,d$ 
listed in Table 1. The two global maxima, labeled $d$, are connected by 
long, almost straight, ridges to the two `inflection maxima', labeled $b$. 
Note that the maxima, at a `height' of $0.36885$, are little higher than 
the inflection maxima, at $0.35556$, so the ridges are almost level. 
This ridge structure is strikingly similar to what is observed in 
experiments \cite{Beysens}. 
Note that the relations $\ln S \propto F\ln t$, and $k_x = u/L_x$ etc., mean 
that the ridges in $S$ become higher, narrower, and closer together as a 
function of ${\bf k}$ with increasing time. We can calculate the angle, 
$\phi$, of the ridges to the $v$-axis (shear direction) from  the slope of 
the line joining points $b$ and $d$ , giving $\tan\phi = 2(1-1/\sqrt{6})$ 
(the same result is obtained from the line joining the saddle points, $c$). 
Using (\ref{cov}), the corresponding angle, $\theta$, in the $(k_x,k_y)$ 
plane is given by $\tan \theta = (\tan \phi)/\gamma t$, i.e.\ the ridges 
tend to align closer to the shear direction as $t$ increases \cite{Beysens}.

For the beam in the $y$ direction, the scattering intensity, $S(k_x,0,k_y,t)$, 
is determined by $F(u,0,w)$. Figure 2 is a contour plot of this function 
with the positions of the stationary points $(u,w)$ indicated. These are a 
minimum (labeled $a$) with $F=0$, at $(0,0)$, a pair of saddle points ($b$), 
with $F=16/45$, at $(\pm 2/\sqrt{3},0)$, a second pair of saddle points ($c$),
with $F=1/4$, at $(0,\pm 1/\sqrt{2})$, and four maxima ($d$), with $F=29/80$, 
at $(\pm 3,\pm 1)/2\sqrt{2}$. Currently available experimental data cannot 
resolve this structure. Instead a broadly elliptical scattering pattern is 
seen in the $(k_x,k_z)$ plane \cite{Beysens}, with the major axis along the 
$k_z$ direction, and the eccentricity increasing in time, as expected from 
the different growth rates (by a factor $t$) of $L_x$ and $L_z$.

\begin{figure}
\narrowtext
\centerline{\epsfxsize\columnwidth\epsfbox{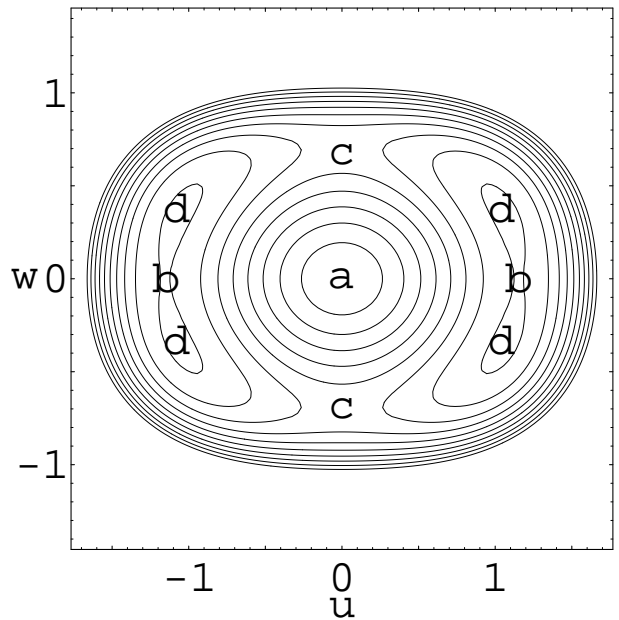}}
\caption{Contour plot of $F(u,0,w)$ showing approximate locations 
of the stationary points $a,b,c,d$ (see text). Contour lines for 
$F < -0.1$ are not shown.} 
\label{fuw}
\end{figure}

Finally, if the beam is parallel to the flow ($k_x=0$), the shear term 
drops out of (\ref{CHbign}). The scattering intensity then has full 
circular symmetry in the $(k_y,k_z)$ plane.

From the asymptotic solution, the expressions derived by Onuki 
\cite{OnukiRev} for the shear-induced contributions to the viscosity, 
$\Delta\eta= -(1/\gamma)\int [d^3k/(2\pi)^3]\,k_xk_y\,S({\bf k},t)$, 
and to the normal stress differences 
$\Delta N_1 = \int [d^3k/(2\pi)^3]\,(k_y^2 - k_x^2)\,S({\bf k},t)$, and 
$\Delta N_2 = \int [d^3k/(2\pi)^3]\,(k_y^2 - k_z^2)\,S({\bf k},t)$ may 
be easily evaluated. Since $S({\bf k},t)$ is a sharply peaked function, 
the factors involving $k_x$, $k_y$, $k_z$ in the integrands can be replaced 
by their values at the peaks ($f$ in Table 1). Using also 
$\int [d^3k/(2\pi)^3]\,S({\bf k},t)=1$ gives, asymptotically,
$\Delta\eta = (b/\gamma^2t^2) \simeq (\ln t/t^3)^{1/2}\gamma^{-2}$, 
$\Delta N_1 = (b/3t) \simeq (\ln t/t)^{1/2}$, and 
$\Delta N_2 = (b/6t) = \Delta N_1/2$.

We now compare our analytical results with the numerical solution of 
(\ref{CHbign}) by Corberi et al.\ \cite{Corberi} in two dimensions 
(the shear-flow, i.e.\  $xy$, plane). Repeating our analysis for 
$d=2$ is straightforward. The dominant stationary points are the 
two maxima labeled $d$ in Table 1 and Figure 1. 
A two-dimensional steepest descent integration over the scaling 
variables $u$ and $v$ yields the self-consistency condition 
\begin{equation}
1 = {\rm const.}\,\frac{\Delta}{\gamma tb} 
\exp\left(\frac{2F_m^{2D}\,b^2}{t}\right)\ ,
\label{key2D}
\end{equation}
instead of (\ref{key}), where (from Table 1) $F_m^{2D} = (37+12\sqrt{6})/180$, 
giving $b(t) = (3t\ln t/4F_m^{2D})^{1/2}$ (to leading logarithmic 
accuracy) instead of (\ref{b}). The structure factor is given by 
$S({\bf k},t) = {\rm const.}\,(\ln A_s)\,A_s^{F^{2D}({\bf q})/F_m^{2D}}$, 
where $A_s = L_xL_y$ is the `scale area', $L_x = \gamma(t^3/b)^{1/2} 
\sim \gamma(t^5/\ln t)^{1/4}$, $L_y = (t/b)^{1/2} \sim (t/\ln t)^{1/4}$, 
${\bf q} = (k_xL_x,k_yL_y)$ is the scaled momentum, and $F_m^{2D}({\bf q})$ 
means $F(u,v,0)$, with $u=k_xL_x$ and $v=k_yL_y$. In other words, the 2D 
structure factor has an identical shape (shown in Figure 1) to the 3D 
structure factor with  $k_z=0$. The main difference is that the numerical 
values of $F_m$ ($0.38333\ldots$) and $F_m^{2D}$ ($0.35555\ldots$) are 
(slightly) different.

In their numerical results, Corberi et al.\ also find two ridge-like 
structures, but the ridges are terminated by two peaks whose relative 
heights {\em oscillate} in time, such that first one peak, then the other 
is the higher. They further speculate that these oscillations 
persist to late times and `characterize the steady state'. Since no 
oscillatory behaviour is found in the asymptotics of our analytical 
solution, we believe that the observed oscillations are slowly-decaying 
preasymptotic transients.

We conclude with some conjectures  about the physically realistic case (for 
binary fluids in the diffusive regime) of a scalar order parameter. These 
are informed by our exact solution for $n=\infty$, and by the way the 
$n=\infty$  solution is known to be  modified for scalar fields in the 
{\em unsheared} case \cite{BH}. First we expect that, for any finite $n$, 
the structure factor will exhibit asymptotic scaling of the form 
$S({\bf k},t) = V_s g({\bf q})$, with $V_s = \Pi_{i=1}L_i$ ($i=x,y,z$) 
and $q_i = k_iL_i$, 
instead of the multiscaling form (\ref{S(k)}). As in the $n=\infty$ case, 
we expect the growth of the characteristic scales for directions normal 
to the flow to obey the same power laws as in the unsheared case, i.e.\ for 
scalar fields, $L_y \sim L_z \sim t^{1/3}$. The growth in the $x$-direction 
can then be deduced from the assumed scaling form for the structure factor: 
if we multiply the two terms on the left-hand side of (\ref{CHbign}) by 
$\phi_{\bf -k}(t)$, and average, the result $L_x \sim \gamma t L_y$ follows 
immediately if we insert the scaling form for $S({\bf k},t)$ and assume 
both terms are of the same order in the scaling limit. This leads to  
the prediction  $L_x \sim \gamma t^{4/3}$.   
The shear-induced viscosity and normal stresses then scale as 
$\Delta\eta \sim (\gamma L_xL_y)^{-1} \sim \gamma^{-2}t^{-5/3}$, and 
$\Delta N_{1,2} \sim 1/L_y^2 \sim t^{-2/3}$.

In the viscous hydrodynamic regime, where without shear $L(t) \sim t$, 
the same heuristics suggest $L_y \sim L_z \sim t$, $L_x \sim \gamma t^2$  
(with corresponding modifications to $\Delta\eta$ and $\Delta N_{1,2}$). 
The predictions for the length scales are consistent with data on polymer  
blends \cite{Lauger}, though it has been suggested \cite{OnukiRev,Hashimoto}
that a stationary state eventually develops for $\gamma t \gg 1$ due to a 
competition between stretching and breaking of domains.

As far as the shape of the structure factor is concerned, we expect 
the scaling function $g({\bf q})$ to be described, in broad terms, by  
$\ln g({\bf q}) \sim F(q_x,q_y,q_z)$, where $F(u,v,w)$ is given by 
(\ref{F}), i.e.\ rather like the large-$n$ theory but without the $\ln V_s 
\sim \ln t$ factor multiplying $F$. In particular, a contour plot of 
$\ln g(q_x,q_y,0)$ should look very similar to Figure 1. (However, the 
behaviour near $q=0$ would be modified by the requirement that $g(0)=0$, 
imposed by the conservation of the order parameter). We hope that these 
predictions will act as a spur to further experimental work.

We thank Caroline Emmott for useful interactions in the early stages of 
this work. This work was supported by EPSRC under grant GR/L97698 (AB), 
and by the Commonwealth Scholarship Commission (NR).

\end{multicols}
\end{document}